\documentclass{appolb}
\usepackage{graphicx}
\usepackage{amssymb}
\usepackage{amsmath}
\usepackage{color}
\begin{document}

\title{Geometrical scaling in high energy collisions \\ and its breaking
\thanks{Presented at the Conference {\em Excited QCD}, Bjelasnica, Sarajevo, Feb. 3 -- 9, 2013.}}
\author{Michal Praszalowicz
\address{M. Smoluchowski Institute of Physics, Jagellonian University, Reymonta 4,
30-059 Krakow, Poland}}

\maketitle

\begin{abstract}
We analyze geometrical scaling (GS) in Deep Inelstic Scattering at HERA and in pp
collisions at the LHC energies and in NA61/SHINE experiment. We argue that GS
is working up to relatively large Bjorken $x \sim 0.1$. This allows to study GS
in negative pion multiplicity $p_{\rm T}$
distributions at NA61/SHINE energies where clear sign of scaling violations
is seen with growing rapidity when one of the colliding partons has Bjorekn
$x \ge 0.1$.
\end{abstract}


\section{Introduction}
\label{intro}

In this short note, following 
Refs.~\cite{McLerran:2010ex}\nocite{Praszalowicz:2011tc,Praszalowicz:2011rm,
Praszalowicz:2012zh}--\cite{Praszalowicz:2013uu} where also an extensive
list of references can be found, we will focus on the scaling law, 
called geometrical scaling
(GS), which has been introduced in the context of DIS \cite{Stasto:2000er}. 
Recently it has been shown that GS is also exhibited by the $p_{\text{T}}$
spectra at the LHC \cite{McLerran:2010ex}--\cite{Praszalowicz:2011rm}. 
An onset of GS in heavy ion collisions at RHIC energies has been reported in 
Ref.~\cite{Praszalowicz:2011rm}.
At low Bjorken $x<x_{\mathrm{max}}$ proton is  
characterized by an intermediate energy scale $Q_{\text{s}}(x)$ --
called saturation scale \cite{sat1,GolecBiernat:1998js} -- defined as the border
line between dense and dilute gluonic systems within a proton (for review
see \emph{e.g.} Refs.~\cite{Mueller:2001fv,McLerran:2010ub}).
For the present study, however, the details of saturation are not of primary
interest, it is the very existence of $Q_{\text{s}}(x)$ which is of  importance.

Here we present analysis of three different pieces of data which exhibit both
emergence and violation of geometrical scaling. In Sect.~\ref{method} we briefly describe
the method used to assess the existence of GS. Secondly, in Sect.~\ref{DIS} we describe
our recent analysis~\cite{Praszalowicz:2012zh} of combined HERA data \cite{HERAdata}
where it has been shown that  GS in DIS 
works very well up to relatively large $x_{\text{max}}\sim0.1$ 
(see also \cite{Caola:2010cy}). Next, in Sect.~\ref{ppLHC}, on the example of the 
CMS $p_{\rm T}$ 
spectra in central rapidity \cite{Khachatryan:2010xs}, 
we show that  GS can be extended to hadronic collisions.  
For particles produced at non-zero rapidities, one
(larger) Bjorken $x=x_{1}$ may leave the domain of GS, \emph{i.e.} $x_{1}>x_{\text{max}}$,
and violation of GS should appear. In Sect.~\ref{ppNA61} we present analysis of very recent
pp data from NA61/SHINE experiment at CERN \cite{NA61} and show that GS is indeed violated once rapidity is increased. We conclude in Sect.~\ref{concl}.

\section{Method of ratios}
\label{method}

Geometrical scaling hypothesis means that some observable $\sigma$ that
in principle depends on two independent kinematical variables, say $x$ and $Q^2$,
in fact depends only on a specific combination of them denoted as  $\tau$:
\begin{equation}
\sigma(x,Q^{2})=F(\tau)/{Q_{0}^{2}}. \label{GSdef}%
\end{equation}
Here function $F$ in Eq.~(\ref{GSdef}) is a dimensionless function of scaling variable
\begin{equation}
\tau=Q^{2}/Q_{\text{s}}^{2}(x).\label{taudef}%
\end{equation}
and
\begin{equation}
Q_{\text{s}}^{2}(x)=Q_{0}^{2}\left(  {x}/{x_{0}}\right)  ^{-\lambda}
\label{Qsat}%
\end{equation}
is the saturation scale. Here $Q_{0}$ and $x_{0}$ are free parameters which can 
be extracted from the data within some specific model for $\sigma$, and exponent 
$\lambda$ is a dynamical quantity of the order of $\lambda\sim0.3$. Throughout
this paper we shall test the hypothesis whether different pieces of data can be described
by formula (\ref{GSdef}) with {\em constant} $\lambda$, and what is the kinematical
range where GS is working satisfactorily. 

In view of Eq.~(\ref{GSdef}) observables $\sigma(x_{i},Q^{2})$ 
for different $x_{i}$'s  should fall on one universal curve, if evaluated not 
in terms of $Q^{2}$ but in terms of $\tau$. This means in turn
that  ratios
\begin{equation}
R_{x_{i},x_{\text{ref}}}(\lambda;\tau_{k})=\frac{\sigma%
(x_{i},\tau(x_{i},Q_{k}^{2};\lambda))}{\sigma(x_{\text{ref}%
},\tau(x_{\text{ref}},Q_{k,\text{ref}}^{2};\lambda))} \label{Rxdef}%
\end{equation}  
should be equal to unity independently of $\tau$. Here for some $x_{\rm ref}$
we pick up all $x_i<x_{\rm ref}$ which have at least two overlapping
points in $Q^2$. 

For $\lambda\neq0$ points of the same $Q^{2}$ but
different $x$'s correspond in general to different $\tau$'s. Therefore one has
to interpolate  $\sigma(x_{\text{ref}},\tau(x_{\text{ref}},Q^{2};\lambda))$ to $Q_{k,\text{ref}%
}^{2}$ such that $\tau(x_{\text{ref}},Q_{k,\text{ref}}^{2};\lambda)=\tau_{k}$.
This procedure is described in detail in 
Refs.~\cite{Praszalowicz:2012zh}.

By tuning $\lambda$ one can make $R_{x_{i},x_{\text{ref}}}(\lambda;\tau
_{k}) \rightarrow 1$ for all $\tau_{k}$.
In order to find optimal value $\lambda_{\rm min}$ that minimizes
deviations of ratios (\ref{Rxdef}) from unity we form the chi-square measure%
\begin{equation}
\chi_{x_{i},x_{\text{ref}}}^{2}(\lambda)=\frac{1}{N_{x_{i},x_{\text{ref}}}%
-1}{\displaystyle\sum\limits_{k\in x_{i}}}\frac{\left(  R_{x_{i}%
,x_{\text{ref}}}(\lambda;\tau_{k})-1\right)  ^{2}}{\Delta R_{x_{i}%
,x_{\text{ref}}}(\lambda;\tau_{k})^{2}} \label{chix1}%
\end{equation}
where the sum over $k$ extends over all points of given $x_{i}$ that have
overlap with $x_{\text{ref}}$, and ${N_{x_{i},x_{\text{ref}}}}$ is a number of
such points.

\section{Deep Inelastic Scattering at HERA}
\label{DIS}

\begin{figure}[ptb]
\centering
\includegraphics[width=8cm,angle=0]{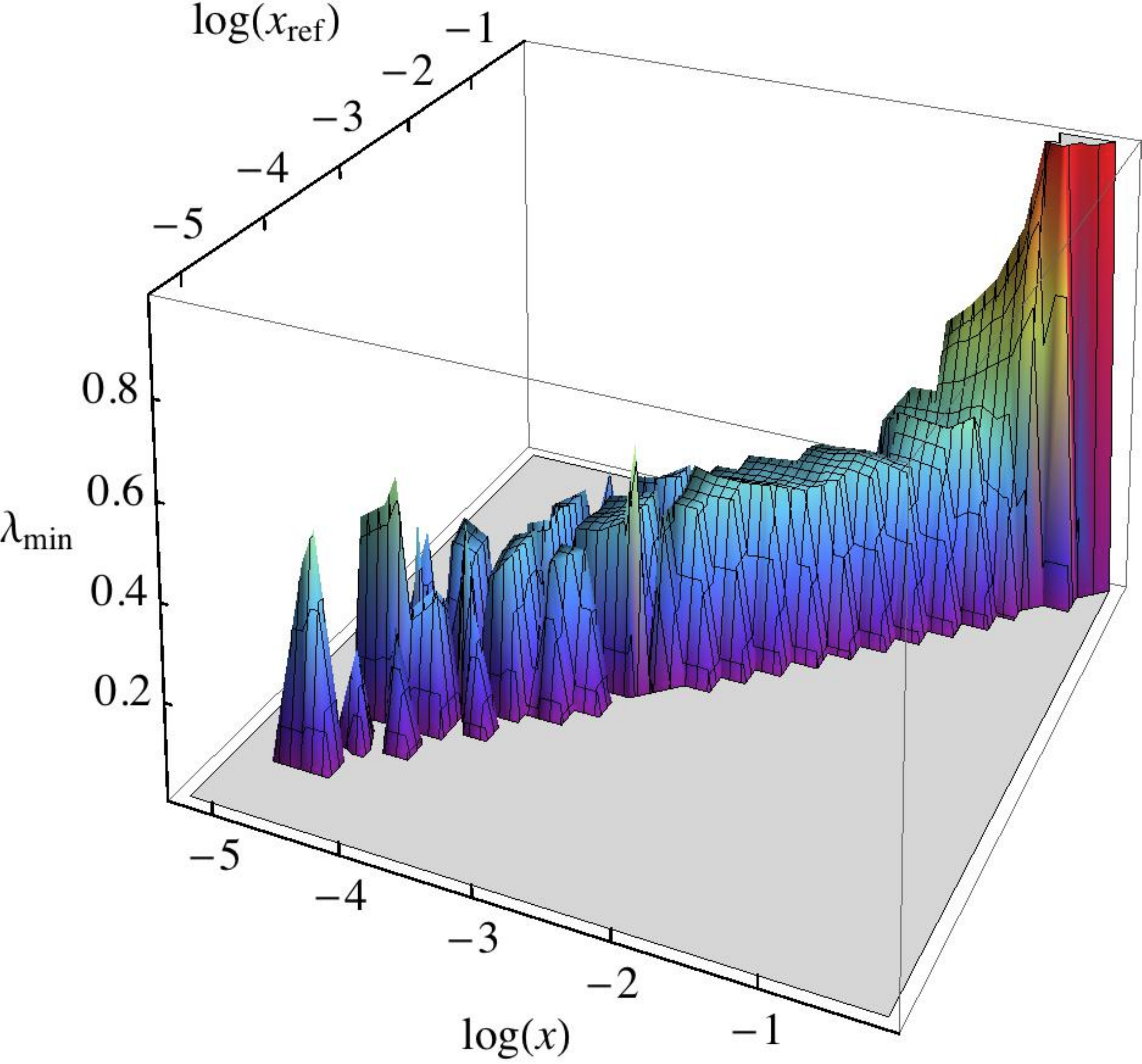}
\caption{Three dimensional
plot of  $\lambda_{\mathrm{min}}(x,x_{\mathrm{ref}})$
obtained by minimization of Eq.~(\ref{chix1}).}%
\label{xlamlog}%
\end{figure}

In the case of DIS the relevant scaling observable is $\gamma^{\ast}p$ cross section
and variable $x$ is simply Bjorken $x$. 
In Fig.~\ref{xlamlog} we
present 3-d plot of $\lambda_{\min}({x,x_{\rm ref}})$ which has been found 
by minimizing  (\ref{chix1}).

Qualitatively,  GS is given by the independence of
$\lambda_{\text{min}}$ on Bjorken $x$ and by the requirement that the
pertinent  value of $\chi_{x,x_{\text{ref}}}^{2}(\lambda_{\text{min}})$ should
be small (for the discussion of the latter see Refs.~\cite{Praszalowicz:2012zh}).  
We see from Fig.~\ref{xlamlog} 
that the stability corner
of $\lambda_{\text{min}}$ 
extends up to $x_{\text{ref}}\lesssim0.1$,
which is well above the original expectations.
In Ref.~\cite{Praszalowicz:2012zh} we have shown that:
\begin{equation}
\lambda = 0.32 - 0.34 \,\,\,\,\, {\rm for} \,\,\,\,\, x \le 0.08.
\end{equation}

\section{Central rapidity $p_{\rm T}$ spectra at the LHC}  
\label{ppLHC}

\begin{figure}[h]
\centering
\includegraphics[scale=0.70]{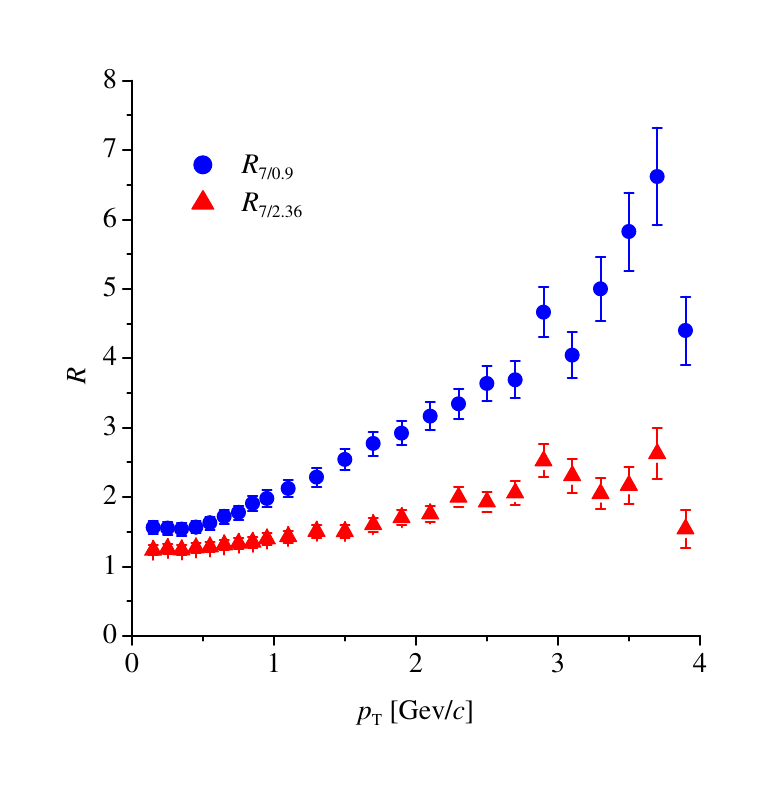} %
\includegraphics[scale=0.70]{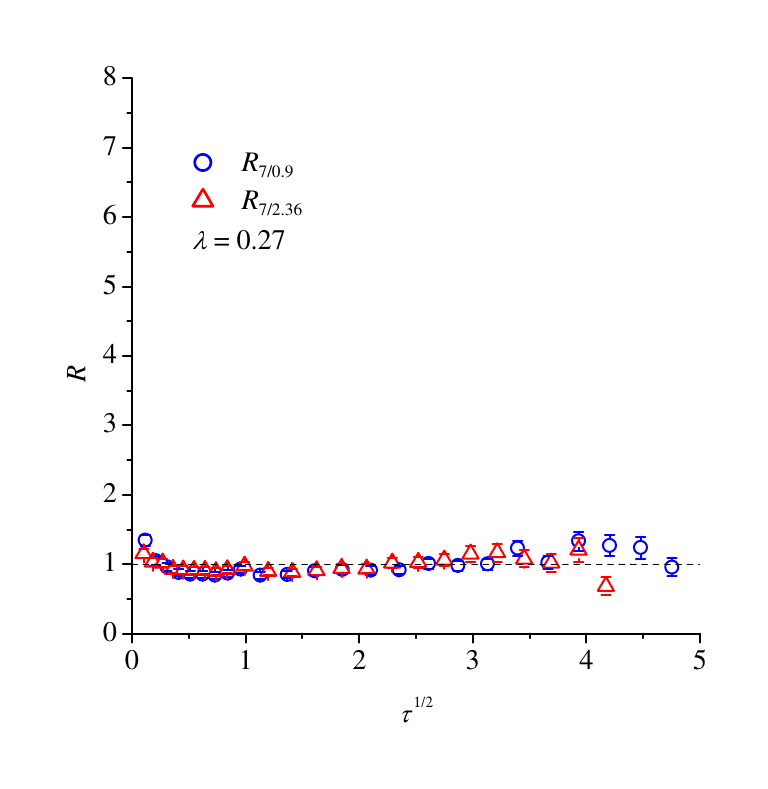}
\caption{Ratios of CMS $p_{\mathrm{T}}$ spectra \protect\cite%
{Khachatryan:2010xs} at 7 TeV to 0.9 (blue circles) and 2.36 TeV (red
triangles) plotted as functions of $p_{\mathrm{T}} $ (left) and scaling
variable $\protect\sqrt{\protect\tau}$ (right) for $\protect\lambda=0.27$. }
\label{ratios1}
\end{figure}

In hadronic collisions at c.m. energy $W=\sqrt{s}$ particles are produced in 
the scattering process of
two patrons carrying Bjorken $x$'s
\begin{equation}
x_{1,2}=e^{\pm y}\,p_{\text{T}}/W.\label{x12}%
\end{equation}
For central rapidities $x=x_1 \sim x_2$. It has been shown that
in this case charged particle multiplicity spectra exhibit GS
\cite{McLerran:2010ex}
\begin{equation}
\left.  \frac{dN}{dy d^{2}p_{\text{T}}}\right\vert _{y\simeq0}=\frac
{1}{Q_{0}^{2}}F(\tau)\label{GSinpp}%
\end{equation}
where $F$ is a universal dimensionless function of the scaling variable
\begin{equation}
\tau=p_{\text{T}}^{2}/Q_{\text{s}}^{2}(x)=
 p_{\text{T}}^{2}/Q_0^2 \,\left( p_{\rm T}/(x_0 \sqrt{s})\right)^\lambda.
\label{taudef}%
\end{equation}
Therfore the scaling observable is  $\sigma(W,p_{\rm T}^2) = {dN}/{dy d^{2}p_{\text{T}}}$
and the method of ratios is applied to the multiplicity distributions at different
energies ($W_i$ taking over the role of $x_i$ in Eq.~(\ref{Rxdef})). For
$W_{\rm ref}$ we take the highest LHC energy of 7~TeV. Therefore one can
form two ratios $R_{W_i,W_{\rm ref}}$ with $W_1 =2.36$ and $W_2 = 0.9$~TeV.
These ratios are plotted in Fig.~\ref{ratios1} for the CMS single non-diffractive spectra
for $\lambda = 0$ and for $\lambda = 0.27$, which minimizes (\ref{chix1})
in this case. We see that original ratios plotted in terms of $p_{\text{T}%
}$ range from 1.5 to 7, whereas plotted in terms of $\sqrt{\tau}$ they are
well concentrated around unity. The optimal exponent $\lambda$ is, however,
smaller than in the case of DIS. Why this so, remains to be understood.

\section{Violation of geometrical scaling in forward rapidity region}
\label{ppNA61}

For $y >0$ two Bjorken $x$'s can be quite different:
$x_{1}>x_{2}$. Therefore looking at the spectra with increasing $y$ one can
eventually reach $x_{1}>x_{\mathrm{max}}$ 
and GS violation should be seen.
To this end we shall use
pp data from NA61/SHINE experiment at CERN \cite{NA61}
at different rapidities $y=0.1-3.5$ and at five different
energies $W_{1,\ldots,5}=17.28,\;12.36,\;8.77,\;7.75$, and $6.28$ GeV.

\begin{figure}[h!]
\centering
\includegraphics[width=6.2cm,angle=0]{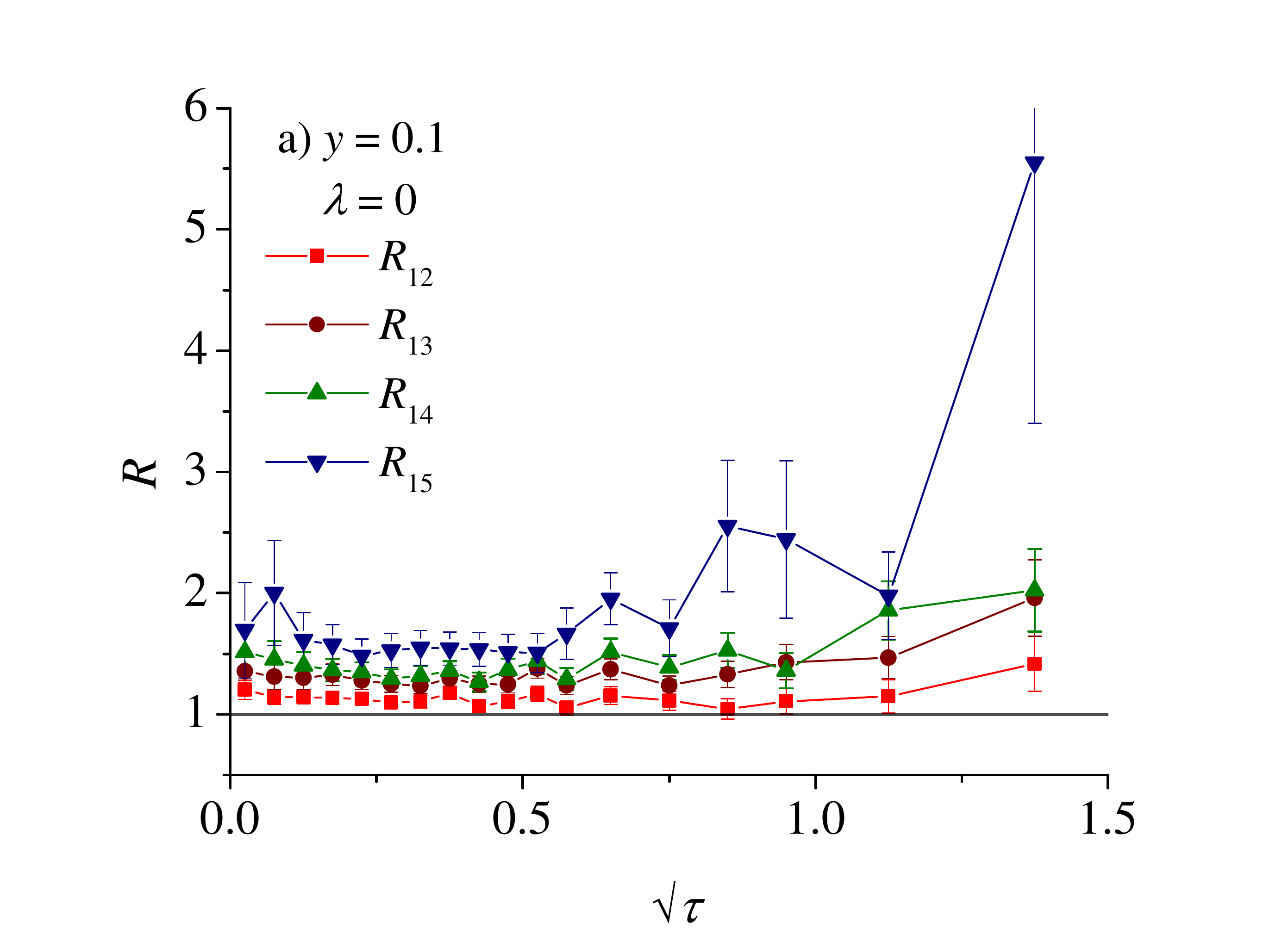}
\includegraphics[width=6.2cm,angle=0]{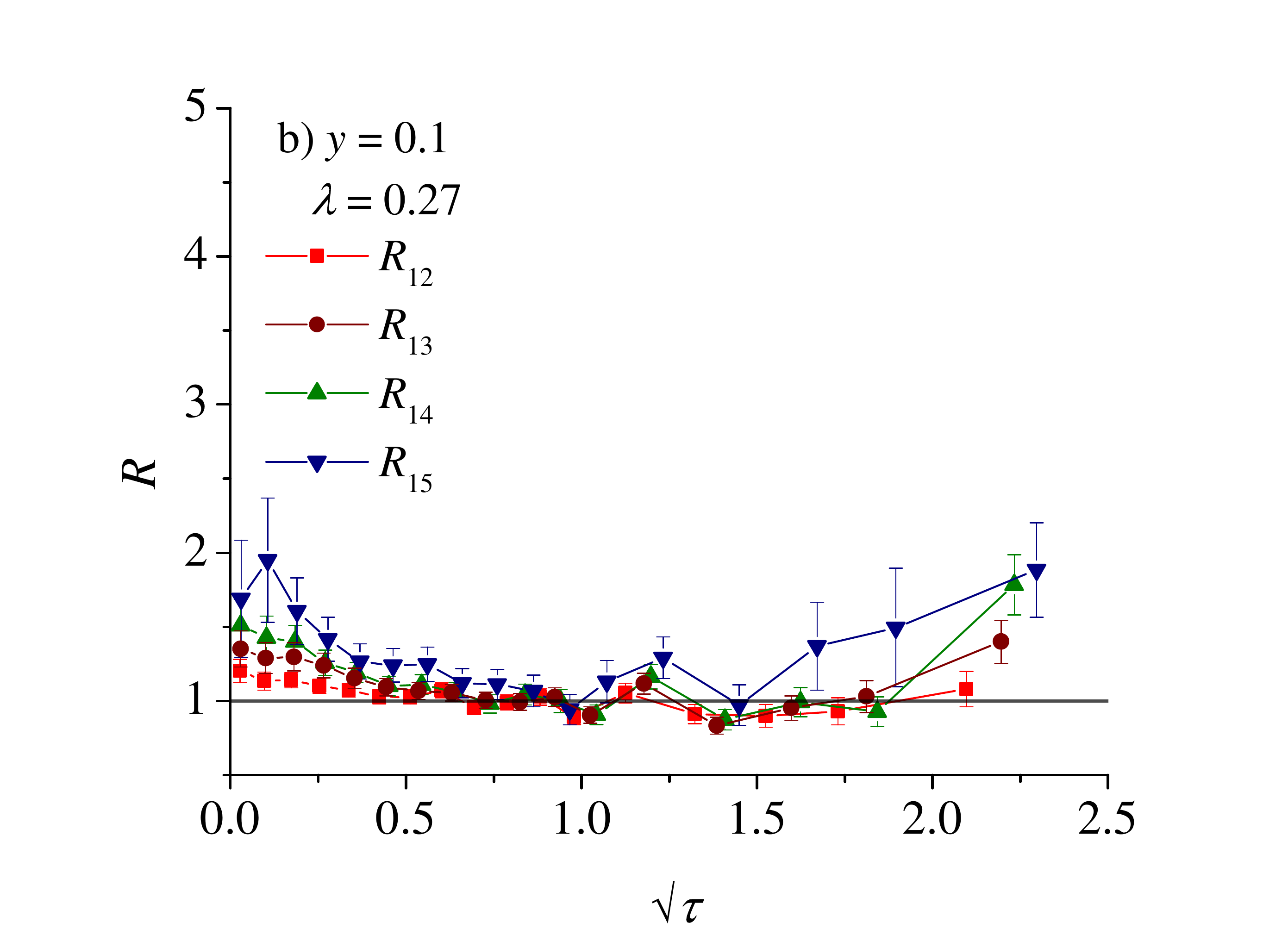} 
\caption{Ratios $R_{1k}$
as functions of $\sqrt{\tau}$ for the lowest rapidity $y=0.1$: a) for
$\lambda=0$ when $\sqrt{\tau}=p_{\mathrm{T}}$ and b) for $\lambda=0.27$ which
corresponds to GS.}%
\label{y01}%
\end{figure}

\begin{figure}[h!]
\centering
\includegraphics[width=6.2cm,angle=0]{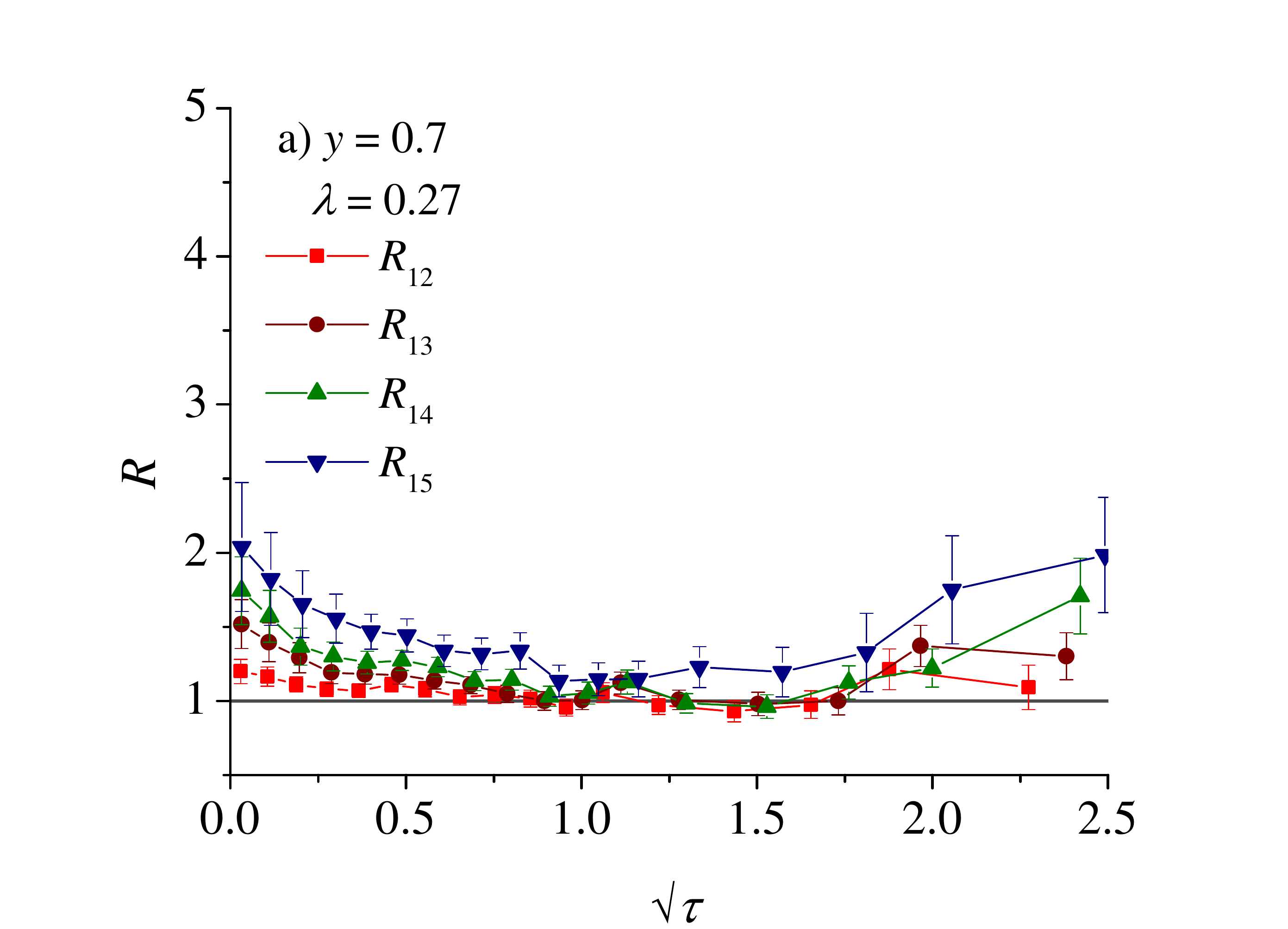}
\includegraphics[width=6.2cm,angle=0]{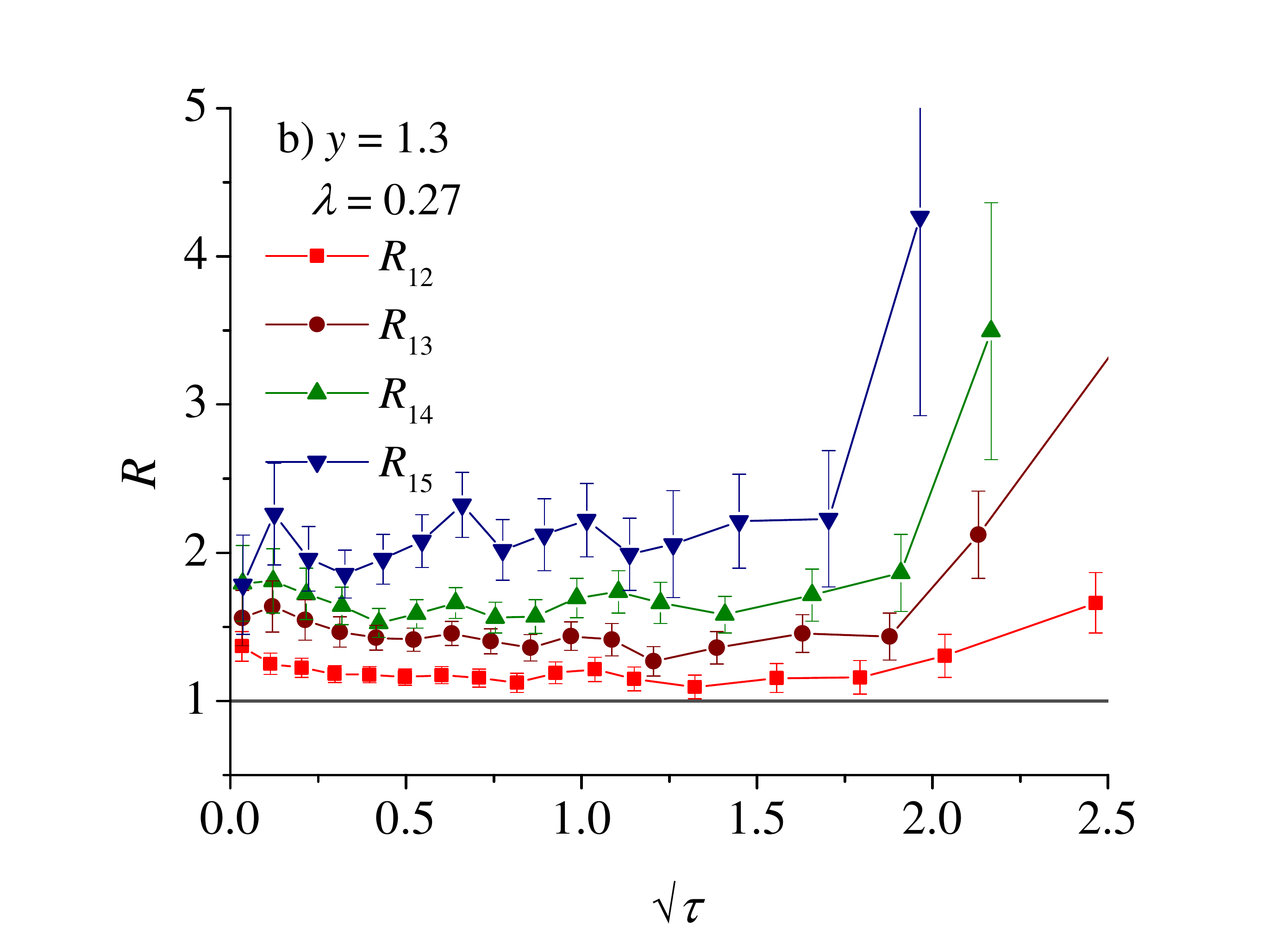} 
\caption{Ratios $R_{1k}$
as functions of $\sqrt{\tau}$ for $\lambda=0.27$ and for different rapidities
 a) $y=0.7$ and b) $y=1.3$. With increase of rapidity,
gradual closure of the GS window can be seen.}%
\label{ys}%
\end{figure}

In Fig.~\ref{y01} we plot ratios $R_{1i}=R_{W_1,W_i}$ (\ref{Rxdef}) for $\pi^-$ spectra in
central rapidity 
for $\lambda=0$ and 0.27. 
For $y=0.1$  the GS region extends towards the smallest energy
because $x_{\rm max}$ is as large as 0.08.
However, the quality of GS is the worst for the lowest energy $W_5$.
 By increasing $y$  some points fall outside  the GS window because  $x_1 \ge x_{\rm max}$, 
 and finally for
$y\ge1.7$ no GS should be present in NA61/SHINE data. This is illustrated 
nicely in Fig.~\ref{ys}.

\section{Conclusions}
\label{concl}

We have shown that GS in DIS works well up to rather large Bjorken $x$'s
with exponent $\lambda = 0.32 - 0.34$.
In pp collisions at the LHC energies in central rapidity GS is seen in 
the charged particle multiplicity spectra, however, $\lambda = 0.27$ in this case. 
By changing rapidity  one can force one of the Bjorken $x$'s of colliding patrons
to exceed $x_{\rm max}$ and GS violation is expected. Such behavior is indeed
observed in the NA61/SHINE pp data.

\bigskip
The author wants to thank M. Ga{z}dzicki and Sz. Pulawski for
the access to the NA61/SHINE data and to T. Stebel for collaboration and remarks. 
Many thanks 
are due to the organizers of this successful series of conferences.
This work was supported by 
the Polish NCN  grant 2011/01/B/ST2/00492.


\begin{thebibliography}{99}
\bibitem{McLerran:2010ex} 
  L.~McLerran and M.~Praszalowicz,
  Acta Phys.\ Polon.\ B {\bf 41} (2010) 1917
  and
  Acta Phys.\ Polon.\ B {\bf 42} (2011) 99.
  
\bibitem{Praszalowicz:2011tc} 
  M.~Praszalowicz,
  Phys.\ Rev.\ Lett.\  {\bf 106} (2011) 142002.
\bibitem{Praszalowicz:2011rm} 
  M.~Praszalowicz,
  Acta Phys.\ Polon.\ B {\bf 42} (2011) 1557
  and
arXiv:1205.4538 [hep-ph].
  

\bibitem{Praszalowicz:2012zh} 
  M.~Praszalowicz  and T.~Stebel,
  JHEP {\bf 1303}, 090 (2013) and 
  arXiv:1302.4227 [hep-ph], to be published in JHEP.
  
  \bibitem{Praszalowicz:2013uu} 
  M.~Praszalowicz,
  arXiv:1301.4647 [hep-ph], to be published in Phys. Rev. {\bf D}.
  
\bibitem{Stasto:2000er} 
  A.~M.~Stasto, K.~J.~Golec-Biernat and J.~Kwiecinski,
  Phys.\ Rev.\ Lett.\  {\bf 86}, 596 (2001).
  

\bibitem{sat1}
L. V. Gribov, E. M. Levin and M. G. Ryskin, Phys. Rept. {\bf 100} (1983) 1; \\
A. H. Mueller and  J-W. Qiu, Nucl. Phys. {\bf 268} (1986) 427; 
A. H. Mueller, Nucl. Phys. {\bf B558} (1999) 285.   

\bibitem{GolecBiernat:1998js} 
  K.~J.~Golec-Biernat and M.~W{\"u}sthoff,
  Phys.\ Rev.\ D {\bf 59} (1998) 014017
  and
  Phys.\ Rev.\ D {\bf 60} (1999) 114023.
  
  \bibitem {Mueller:2001fv}A.~H.~Mueller, \emph{Parton Saturation: An Overview},
arXiv:hep-ph/0111244.

\bibitem {McLerran:2010ub}L.~McLerran, 
{Acta Phys.\ Pol.\ B} \textbf{41}, 2799 (2010).  

\bibitem {HERAdata}C.~Adloff \textit{et al.} [H1 Collaboration],
{Eur.\ Phys.\ J.\ C} \textbf{21} (2001) 33; 
S.~Chekanov \textit{et al.} [ZEUS Collaboration],
{Eur.\ Phys.\ J.\ C} \textbf{21} (2001) 443. 

\bibitem{Caola:2010cy} 
  F.~Caola, S.~Forte and J.~Rojo,
  Nucl.\ Phys.\ A {\bf 854}, 32 (2011).
  

\bibitem{Khachatryan:2010xs}  V.~Khachatryan \textit{et al.} [CMS
Collaboration], 
JHEP \textbf{1002} (2010) 041 
and 
Phys.\ Rev.\ Lett.\ \textbf{105} (2010) 022002
and 
JHEP \textbf{1101} (2011) 079. 
%


  \bibitem{NA61}
 N. Abgrall {\em et al.} [NA61/SHINE Collaboration],
 {\em Report from the NA61/SHINE experiment
at the CERN SPS} CERN-SPSC-2012-029, SPSC-SR-107;\\ 
 A. Aduszkiewicz, Ph.D. Thesis in prepartation, University of Warsaw, 2013;\\
 Sz. Pulawski, talk at 9th Polish Workshop on Relativistic Heavy-Ion Collisions,
Krak{\'o}w, November 2012
 and private communication.
 
\end{thebibliography}
\end{document}